\documentclass[12pt,a4paper,reqno]{article}

\usepackage{amsmath,latexsym,amssymb,amsthm}
\usepackage{graphicx}

\usepackage{amsmath}\usepackage{amssymb}
\usepackage{amsthm}

\newcommand{\semilinespace}{\vspace{0.5\baselineskip}}

\newcommand{\be}{\begin{equation}}
\newcommand{\ee}{\end{equation}}
\renewcommand{\(}{\left(}
\renewcommand{\)}{\right)}

\newcommand{\e}{\mathrm{e}}

\renewcommand{\H}{\mathcal{H}}

\newcommand{\ox}{\otimes}

\newcommand{\<}{\langle}
\renewcommand{\>}{\rangle}

\theoremstyle{definition}

\theoremstyle{remark}

\numberwithin{equation}{section}

\begin{document}

\title{\bf{Time, chance and quantum theory}}

\author{Anthony Sudbery$^1$\\[10pt] \small Department of Mathematics,
University of York, \\[-2pt] \small Heslington, York, England YO10 5DD\\
\small $^1$ as2@york.ac.uk}

\date{20 October 2015}

\maketitle

\begin{abstract}

I propose an understanding of Everett and Wheeler's relative-state interpretation of quantum mechanics, which restores the feature of indeterminism to the theory. This incorporates a theory of probability as truth values in a many-valued logic for future statements, and a contextual theory of truth which gives objective and subjective perspectives equal validity.

\end{abstract}

\section{Introduction}

When quantum mechanics emerged in the 1920s as the basic format in which all subsequent theories of physics would be couched, it exhibited two distinct features which represented its radical conceptual departure from classical physics. These two features have been given \cite{QMPN} the confusingly similar labels
of \emph{indeterminacy} and \emph{indeterminism}.

\emph{Indeterminacy} is the more startling and harder to grasp of these two new concepts, yet it has the clearer mathematical formulation in the theory. Physically, it is the idea that physical quantities need not have definite values. Mathematically, this is expressed by the representation of such quantities by hermitian operators instead of real numbers, and the mathemtical fact that not every vector is an eigenvector of such an operator. This yields the physical notion of \emph{superposition}, well known as a stumbling block in understanding the theory.

\emph{Indeterminism}, on the other hand, is simply the opposite of determinism, as canonically expressed by Laplace \cite{Laplace:determinism}. It is the denial of the idea that the future of the universe is rigidly determined by its present state. It can be seen as not so much a rejection, more a reluctant abandonment: if Laplace's determinism is seen, not as a statement of faith, but as a declaration of intent --- ``We will look for an exact cause in the past of everything that happens in the physical world" --- then quantum theory is an admission of failure: ``We will not try to find the reason why a radioactive nucleus decays at one time rather than another; we do not believe that there is any physical reason for its decay at a particular time."

Indeterminism, unlike indeterminacy, does not present any conceptual difficulty. In the history of human thought, the determinism of classical physics is a fairly recent and, to many, a rather troubling idea. To the pre-scientific mind, whether in historical times or in our own childhood, the apparent fact of experience is surely that there are things which might happen but they might not: you can never be sure of the future. As it is put in the Bible, ``time and chance happeneth to them all".

On simply being told that the new theory of physics is to be indeterministic, one might think one could see the general form that theory would take: given a mathematical description of the state of a physical system at time $t_0$, there would be a mathematical procedure which, instead of yielding a single state at each subsequent time $t > t_0$, would yield a set of possible states and a probability distribution over this set. This would give a precise expression of the intuition that some things are more likely to happen than others; the probabilities would be objective facts about the world.

But this is not how indeterminism is manifested in quantum mechanics. As it was conceived by Bohr and formalised by Dirac \cite{Dirac:book} and von Neumann \cite{vonNeumann:QM} and most textbooks ever since (the ``Copenhagen interpretation"), the indeterminism does not enter into the way the world develops by itself, but only in the action of human beings doing experiments. In the basic laws governing physical evolution, probabilities occur only in the projection postulate, which refers to the results of experiments. The idea that experiments should play a basic part in the theory has been trenchantly criticised by John Bell \cite{Bell:piddling}. This unsatisfactory, poorly defined and, frankly, ugly aspect of the theory consititutes the \emph{measurement problem} of quantum mechanics.

It could be argued that the indeterminism of quantum theory is played down in the Copenhagen interpretation, in the sense that the role of time in physical probability is minimised. This is not a formal feature of the theory, but an aspect of its presentation. Experiments are presented as if they took no time to be completed; probabilities are properties of the instantaneous state. Heisenberg referred to the contents of the state vector as representing potentiality, and this aspect of the state vector has been emphasised by de Ronde \cite{deRonde:potentiality}, but it seems to me that the idea of something being potential looks to the future for its actualisation.

The theory remained in this state (and some say it still does) until the papers of Everett and Wheeler in 1957. Everett, as endorsed by Wheeler, argued that there was no need for the projection postulate and no need for experiments to play a special role in the theory. But this did not give a satisfactory place for indeterminism in the theory: on the contrary, quantum mechanics according to Everett is essentially deterministic and probability is generally regarded as a \emph{problem} for the Everett-Wheeler interpretation.

There are at least four different concepts that go under the name of ``probability" (or, if you think that it is a single concept, four theories of probability) \cite{Gillies}. Probability can be a strength of logical entailment, a frequency, a degree of belief, or a propensity. Physicists tend to favour what they see as a no-nonsense definition of probability in terms of frequencies, but in the context of physics this is ultimately incoherent. Classical deterministic physics has a place for probability in situations of incomplete knowledge; this is how Laplace considers probability. The concept being used here is degree of belief. Some quantum physicists \cite{QBism} would like to retain this concept in the indeterministic realm. Many, however, would reject this subjectivist approach as abandoning the ideal of an objective theory. The most appropriate concept for an indeterministic theory, as outlined above, would seem to be that of propensity; if there is a set of possible future states for a system, the system has various propensities to fall into these states. This gives an objective meaning of ``probability". This concept is also appropriate to the Copenhagen interpretation of quantum mechanics: an experimental setup has propensities to give the different possible results of the experiment. The Everett-Wheeler picture, however, seems to have no place for propensities.  

Everett liked to describe the universal state vector, which occupies centre stage in his theory, as giving a picture of \emph{many worlds}; Wheeler, I think rightly, rejected this language. Both views, however, attribute simultaneous existence to the different components of the universal state vector which are the seat of probabilities in the Copenhagen interpretation. It is hard to see how propensities can attach to them. Everett, consequently, pursued the idea of probabilities as frequencies, and even claimed to derive the quantum-mechanical formula for probability (the Born rule) rather than simply postulating it. It is generally agreed, however, that his argument was circular (\cite{Wallace:multiverse}, p. 127).

In this article I will explore the Everett-Wheeler formulation of quantum mechanics from the point of view of a sentient physical system inside the world, such as each of us. I argue that it is necessary to recognise the validity of two contexts for physical propositions: an external context, in which the universal state vector provides the truth about the world and its development is deterministic; and an internal context in which the world is seen from a component of the universal state vector. In the latter context the development in time is indeterministic; probabilities apply, in a particular component at a particular time, to statements referring to the future. This leads to a fifth concept of probability: the probability of a statement in the future tense is the \emph{degree of truth} of that statement. I conclude with a formulation of the predictions of quantum mechanics from the internal perspective which fits the template for an indeterministic theory described earlier in this section: for each state at time $t_0$, the universal state vector provides a set of possible states at subsequent times $t > t_0$ and a probabiilty distribution over this set.

I start by arguing for the relative-state interpretation of Everett and Wheeler as the best approach to the problem of understanding quantum theory.

\section{The relative-state understanding of quantum mechanics}

How can we approach the problem of understanding quantum mechanics?

First, some principles: what understanding a scientific theory means to me. I want to take our best science seriously; whatever a successful theory supposes, I am disposed to believe that that is the truth about the world. But secondly, I want to take my own experience seriously. Too often, expositors of science tell us that what we experience is an illusion. Of course, illusions do exist: when I see a stick bending as it is put into water at less than a right angle to the surface of the water, that is an illusion, and it can be demonstrated to be an illusion by immediate appeal to other experiences.\footnote{Such illusions are less common than is often thought. We are told that Copernicus discovered that it was an illusion that the sun goes round the earth. Not so: it is \emph{true} that the sun goes round the earth. Einstein taught us that any such statement is relative to a frame of reference, and in my rest frame the sun does indeed go round me once a day. This reflection is relevant to the understanding of quantum mechanics suggested here, with ``frame of reference" replaced by ``perspective".} But there are other experiences that are too basic to be illusory. Such fundamental aspects of our experience as time, and consciousness, and free will, and the solidity of solids \cite{Stebbing}, cannot be illusions. It may be that we have a false theory of these phenomena -- for example, that free will consists of an interruption to the laws of nature, or that the solid state consists of a mathematical continuum -- it may, indeed, be difficult to define them precisely -- but that does not mean that the phenomena themselves are not real.

Now here's the problem. It seems to be impossible to apply both of these principles in understanding quantum mechanics. One of the basic facts of our experience which I cannot disbelieve is that (properly conducted) experiments have well-defined, unique results. But there are no such unique results of experiment in pure quantum theory. 

What do I mean by ``pure quantum theory"? The mathematical machinery is not in question: Hilbert spaces to describe the possible states of systems, tensor products to combine them, Hamiltonians to describe interactions, the Schr\"odinger equation to govern evolution. That is clear-cut. But if the theory is to include our unique experience, it seems that further, more vague, elements have to be added to this clear theory. 
What is traditionally taught as quantum mechanics therefore includes, in addition to the Schr\"odinger equation, the ``collapse postulate": a stochastic evolution which delivers a state vector incorporating the uniqueness that we experience. For some reason, this only happens after a ``measurement", though nobody has ever made precise what exactly a measurement is; so it does not describe our experience when we are not making measurements \cite{verdammte}. I therefore follow Everett and Wheeler in not including this part of the text-book account in what I mean by ``quantum theory". 

The preceding criticisms do not apply to theories of spontaneous collapse like the GRW theory \cite{GRW}, which was designed to overcome these objections, but such theories go beyond orthodox quantum theory and, in principle, are empirically distinguishable from it. This response to the challenge of understanding quantum theory essentially consists of saying ``It's hopeless --- quantum mechanics can't be understood; we need to replace it with a different theory". That may be correct, but in this article  I want to persevere in the attempt to understand the pure theory of quantum mechanics.

It then seems that my two basic principles are incompatible. If the science of quantum theory is true, then experiments have no unique results and our experience of such uniquenesss cannot be trusted. We cannot take quantum mechanics seriously and at the same time take our basic experience seriously. 

This dilemma has the same form as some perennial philosophical problems. For example,
\begin{enumerate}
\item The existence of space-time \emph{vs} the passage of time;
\item Determinism \emph{vs} free will (or, indeed, \emph{in}determinism \emph{vs} free will);
\item The physical description of brain states \emph{vs} conscious experience;
\item Duty \emph{vs} self-interest
\end{enumerate}
This general class of philosophical problems has been discussed by Thomas Nagel in his book \emph{The View from Nowhere} \cite{Nagel:nowhere}. Each of the above examples is, or seems to be, a contradiction between two statements or principles, both of which we seem to have good reason to believe. In each case one of the statements is a general universal statement --- what Nagel calls ``a view from nowhere" --- to which assent seems to be compelled by scientific investigation or moral reflection; the other is a matter of immediate experience, seen from inside the universe (a view from ``now here").

Nagel suggests that we can resolve these contradictions by recognising that the apparently opposed statements can both be true, but in different contexts (or from different perspectives). Statements in the view from nowhere do not contradict statements in the view from now here, they just do not engage with them; the two types of statement are \emph{incommensurable}. But, provided we are careful to specify the context of each statement, we will see that in each pair, the apparently opposed statements are \emph{compatible}.

In order to see how these ideas apply to quantum mechanics, let us look at the famous example of Schr\"odinger's cat.  

Schr\"odinger's sad story \cite{Schrcat} is often presented as a challenge to quantum mechanics. When the unfortunate cat has been in 
Schr\"odinger's diabolical device for a time $t$, the crude argument (not Schr\"odinger's!) goes, its state is
\[
|\psi_\text{cat}(t)\> = \e^{-\gamma t}|\text{alive}\> + \sqrt{1 - \e^{-2\gamma t}}|\text{dead}\>
\] 
So why don't we see such superpositions of live and dead cats?

The answer is simple. If we are watching the cat, hoping to see a superposition like the above, the interaction by which we see it actually produces the entangled state
\[ 
|\Psi(t)\> = \e^{-\gamma t}|\text{alive}\>_\text{cat}|\stackrel{\centerdot\;\centerdot}{\smile}\>_\text{observer} + \sqrt{1 - \e^{-2\gamma t}}|\text{dead}\>_\text{cat}|\stackrel{\centerdot\;\centerdot}{\frown}\>_\text{observer}
\]
in which $|\stackrel{\centerdot\;\centerdot}{\smile}\>$ is the observer state of seeing a live cat and $|\stackrel{\centerdot\;\centerdot}{\frown}\>$ is the state of seeing a dead cat. Nowhere in this total state is there an observer seeing a superposition of a live and a dead cat.

But that doesn't tell us what there actually \emph{is} in the state $|\Psi(t)\>$. In order to understand the meaning of this superposition, let us look at it more carefully.

If the observer is watching the cat continuously over the period from time $0$ to time $t$, they will be able to note the time, if any, at which they see the cat die. Then the joint state of the cat and the observer is something like
\begin{multline}
|\Psi(t)\> = \e^{-\gamma t}|\text{alive}\>_\text{cat}|\text{``The cat is alive''}\>_\text{observer}\\ + \sqrt{2\gamma}\int_0^{t'}\e^{-\gamma t'}|\text{dead}\>_\text{cat}
|\text{``I saw the cat die at time $t'$''}\>_\text{observer}dt'
\end{multline}
in which the observer states contain propositions which are physically encoded in the brain of the observer. But what is their status as propositions; are they true or false?  Each is believed by a brain which has observed the fact it describes, and that fact belongs to reality. As a
human belief, each statement could not be more true. Yet they cannot all be true, for they contradict each other.

This conflict shows the necessity of considering the perspective from which a statement is made when discussing its truth value. When this is done, it becomes possible for contradictory statements to be simultaneously true, each in its own context. 

In general, the state of the universe can be expanded in terms of the states of any observer inside the universe as
\[
|\Psi(t)\> = \sum_n|\eta_n\>|\Phi_n(t)\>
\]
where the $|\eta_n\>$ form an orthonormal basis of observer states, which we can take to be eigenstates of definite experience; the $|\Phi_n(t)\>$ are the corresponding states of the rest of the universe at time $t$. The latter are not normalised; indeed, most of them will be zero. It is only possible for the observer to experience being in one of the states $|\eta_n\>$, and in this state it is true for the observer that the only experience they have is $\eta_n$; the observer is justified, at time $t$, in deducing that the rest of the universe is in the unique state $|\Phi_n(t)\>$. This is the \emph{internal} truth relative to the experience state $|\eta_n\>$.

But there is also the \emph{external} truth that the state of the whole universe is $|\Psi(t)\>$. From this standpoint all the experiences $\eta_n$ truly occur. In Everett's terms, $|\Phi_n(t)\>$ is the \emph{relative} state of the rest of the universe relative to the observer's state $|\eta_n\>$.

Thus there are the following two types of truth involved.

{\bf External truth:} The truth about the universe is given by a state vector $|\Psi(t)\>$ in a Hilbert space $\mathcal{H}_U$, evolving according to the Schr\"odinger equation. If the Hilbert space can be factorised as 
\[
\mathcal{H}_U = \mathcal{H}_S\ox\mathcal{H}_E
\]
where $\mathcal{H}_S$ contains states of an experiencing observer, then 
\[
|\Psi(t)\> = \sum_n|\eta_n\>|\Phi_n(t)\>
\]
and all the states $|\eta_n\>$ for which $|\Phi_n(t)\> \ne 0$ describe experiences which actually occur at time $t$.

\medskip

{\bf Internal truth} from the perspective $|\eta_n\>$: I actually have experience $\eta_n$, which tells me that the rest of the universe is in the state $|\Phi_n(t)\>$. This is an objective fact; everybody I have talked to agrees with me.

\medskip

This distinction between internal and external truth is an example of Nagel's distinction between the view from nowhere and the view from ``now here".  Scientists might be tempted to exalt the external statement as the objective truth, downgrading internal statements as merely subjective. Indeed, Nagel himself uses the terminology of ``objective'' and ``subjective'' \cite{Nagel:subjobj}. But he does not use a dismissive qualifier like ``merely'' to denigrate the subjective: he is at pains to emphasise that the truth of an internal statement has a vividness and immediacy, resulting from the fact that it is actually experienced, compared to which external truth is ``bleached-out''. This applies most obviously in contexts like ethics and aesthetics, but we would do well to remember it in our scientific context; as I have pointed out above, it is the internal statement which has the scientific justification of being supported by evidence, and is objective in the usual sense that it is empirical and is agreed by all observers who can communicate with each other.

But the situation is more complicated than this might suggest. It is not that there is a God-like being who can survey the whole universe and make statements about the universal state vector, distinct from us physical beings who are trapped in one component of $|\Psi\>$. We physical beings are the ones who make statements about $|\Psi\>$, for good theoretical reasons, from our situation in which we experience just the one component $|\eta_n\>|\Phi_n\>$. From that perspective, what are we to make of the other components $|\eta_m\>|\Phi_m\>$?

Consider a measurement process, in which an initial state $|\Psi(0)\> = $ $|\eta_0\>|\Phi_0(0)\>$, containing only one experience $\eta_0$, develops in time $t$ to an entangled state $|\Psi(t)\> = \sum|\eta_n\>|\Phi_n(t)\>$. The external statement is: 
\begin{quotation} $|\Psi(t)\>$ represents a \emph{true} statement about the universe, and all its components are \emph{real}. \end{quotation}
The observer who experiences only $|\eta_n\>$ must say: 
\begin{quotation} I know that only $|\eta_n\>$ is {\bf real} (because I experience only that), and therefore $|\Phi_n(t)\>$ represents a {\bf true} statement about the rest of the universe. But I also know that $|\Psi(t)\>$ is \emph{true} (because I've calculated it). The other $|\eta_m\>|\Phi_m\>$ represent things that {\bf might have happened} but {\bf didn't}.
\end{quotation}
These statements are font-coded, using bold type for internal (vivid, experienced) judgements, and italic for external (pale, theoretical) ones, even though these are made by an internal observer.

It is a constant temptation in physics, on finding a quantum system in a superposition $|\phi\> + |\psi\>$, to think that it is either in the state $|\phi\>$ or in the state $|\psi\>$. This, after all, is the upshot when we look at the result of an experiment. Despite the stern warnings of our lecturers when we are learning the subject, and the proof from the two-slit experiment that ``$+$'' cannot mean ``or'' (\cite{QMPN}, pp. 188-9), we all probably slip into this way of thinking at times; and the common-sense view of Schr\"odinger's cat seems to justify it. On the other hand, the many-worlds view insists that both terms in the superposition are real, and therefore ``$+$'' means ``and''. What I am suggesting here is that both ``and'' and ``or'' are valid interpretations of ``$+$'' in different contexts: ``and'' in the external view, ``or'' in the internal view. 

\semilinespace

\noindent {\bf How Many Worlds?} The foregoing needs some refinement. Quantum superposition is not just a single binary operation ``$+$'' on states, but is modified by coefficients: $a|\phi\> + b|\psi\>$ is a weighted superposition of the (normalised) states $|\phi\>$ and $|\psi\>$. In interpreting ``$+$'' as ``or'', it is easy to incorporate this weighting of the disjuncts by interpreting it in terms of probability. But if we interpret ``$+$'' as ``and'', as in the many-worlds interpretation, what can it mean to weight the conjuncts? Let us look back to the paradigmatic geometrical meaning of vector addition. ``Going north-east'' is the vector sum of ``going north'' and ``going east'', and does indeed mean going north \emph{and} going east at the same time. But ``going NNE'' also means going north and going east at the same time; only there is more going north than going east. So if a superposition $a|\phi\> + b|\psi\>$ of macroscopic states $|\phi\>$ and $|\psi\>$ means that both $|\phi\>$ and $\psi\>$ are real in different worlds, we must accept that they are not ``equally real'', as is often carelessly stated by Everettians (including Everett himself \cite{Everett} --- in a footnote --- but not Wheeler \cite{Wheeler}), but that they are real to different extents $|a|^2$ and $|b|^2$. Adding up these degrees of reality, we then find that there are not many worlds but ($|a|^2 + |b|^2 =$) one.

Another argument for this conclusion uses the physical observable of particle number. The many-worlds view regards the state vector $|\Psi(t)\> = \sum_n|\eta_n\>|\Phi_n(t)\>$ as describing many (say $N$) worlds, with $N$ different copies of the observer having the different experiences $|\eta_n\>$. Suppose the observer's name is Alice. There is an observable called Alice number, of which each of the states $|\eta_n\>|\Phi_n(t)\>$ is an eigenstate with eigenvalue 1. Then $|\Psi(t)\>$ is also an eigenstate of Alice number with eigenvalue 1 (not $N$). There is only one Alice.

\section{Probability and the Future}

\subsection{Probability}

The observer in this measurement process might go on to say:
\begin{quotation}
I \textbf{saw} a transition from $|\eta_1\>$ to $|\eta_n\>$ at some time $t' < t$. But I \emph{know} that $|\Psi(t')\>$ didn't collapse. The other $|\eta_m\>|\Phi_m\>$ \textbf{might} come back and interfere with me in the future; but this has very low \textbf{probability}.
\end{quotation}
This shows how the conflicting statements about time development in the Everett-Wheeler and Copenhagen interpretations can after all be compatible. The continuous Schr\"odinger-equation evolution postulated by the Everett-Wheeler interpretation refers to the external view; the collapses postulated by the Copenhagen interpretation refer to the internal view, i.e.\ to what we actually see. The occurrence of collapse, in the form of perceived quantum jumps in the memory of an observer, becomes a theorem rather than a postulate (see \cite{verdammte} for some indications, but work remains to be done on this). We also see that it is only the internal statement that mentions probability. But what does it mean?

The meaning of probability is a long-standing philosophical problem (see, for example, \cite{Gillies}). There are in fact several distinct concepts which go by the name of probability, sharing only the fact that they obey the same mathematical axioms. The clearest of these, perhaps, is degree of belief, which has the advantage that it can be defined operationally:\ someone's degree of belief in a proposition is equal to the odds that they are prepared to offer in a bet that the proposition is true. The subjective nature of this concept seems to chime with the fact that it belongs in internal statements, as we have just seen, and indeed similar views of probability are often adopted by Everettians even though their general stance is objectivist.

However, we have also seen that ``internal'' should not be equated with ``subjective'', and our experience in a quantum-mechanical world seems to require a description in terms of objective chance. Things happen randomly, but with definite probabilities that cannot be reduced to our beliefs. The value of the half-life of uranium 238 is a fact about the world, not a mere consequence of someone's belief. 

Such objective probability (or ``chances") can only refer to future events. 

\subsection{The Future}

What kinds of statements can be made at time $t_0$ about some future time $t > t_0$, if the universal state vector is known to be $|\Psi(t_0)\>$ and its decomposition with respect to experience states of a particular observer is $\sum_n|\eta_n\>|\Phi_n(t_0)\>$? From the external perspective, the future state $|\Psi(t)\>$ is determined by the Schr\"odinger equation and there is no question of any probability. From the internal perspective relative to an experience state $|\eta_n\>$, there is a range of possible future states $|\eta_m\>$, and probabilities must enter into the statement of what the future state will be. But here is a fundamental problem: there is \emph{no such thing} as what the future state will be. As Bell pointed out \cite{Bell:cosmologists}, quantum mechanics gives no connection between a component of $|\Psi\>$ at one time and any component at another time; so what is it that we can assign probabilities to? How can ``the probability that my state will be $|\eta_m\>$ tomorrow" mean anything when ``my state will be $|\eta_m\>$ tomorrow" has no meaning?

This puzzle takes us back to ways of thinking that are much older than quantum mechanics, indeed older than all of modern science. The success of Newtonian deterministic physics has led us to assume that there always is a definite future, and even when we drop determinism we tend to continue in the same assumption. There is a future, even if we do not and cannot know what it will be. But this was not what Aristotle believed, and maybe it is not what we believed when we were children. 

Aristotle, in a famous passage \cite{seabattle}, considered the proposition ``There will be a sea-battle tomorrow''. He argued that this proposition is neither true nor false (otherwise we are forced into fatalism). Thus he rejected the law of excluded middle for future-tense statements, implying that they obey a many-valued logic. Modern logicians \cite{Prior:logic} have considered the possibility of a third truth-value in addition to ``true" or ``false", namely $u$ for ``undetermined", for future-tense statements. But, interestingly, Aristotle admitted that the sea-battle might be more or less likely to take place. This suggests that the additional truth values needed for future-tense statements are not limited to one, $u$, but can be any real number between 0 and 1 and should be identified with the probability that the statement will come true.\footnote{This idea motivated \L ukasiewicz \cite{Luk:indeterminism} in formulating modern many-valued logic, and has been applied to quantum mechanics by Pykacz \cite{Pykacz}.} Turning this round gives us an objective form of probability which applies to future events, or to propositions in the future tense; in a slogan,
\[
\text{Probability} = \text{degree of future truth}.
\]

A form of temporal logic incorporating this idea is developed in \cite{logicfuture}. It contains a lattice of propositions in the context of a particular observer and a particular time $t_0$ (``now"). The lattice is generated by propositions corresponding to the observer's experience eigenstates, labelled by time $t$. The observer is assumed to have a memory of experiences occurring before $t_0$:\ thus propositions in the present tense ($t = t_0$) and in the past tense ($t < t_0$) have truth values 0 or 1, as in classical logic, but those in the future tense ($t > t_0$) can take any truth value in the real interval $[0,1]$. General propositions are formed from these dated experience propositions by means of conjunction, disjunction and negation (``and", ``or" and ``not"). A conjunction of experience propositions is a history in the sense of the consistent-histories formulation of quantum mechanics; then a conjunction of histories is another history, but the general proposition is a disjunction of histories -- something which is not usually considered in this formulation. It is shown in \cite{logicfuture} that, given a weak form of the assumption that the histories are consistent, the usual formula for the probability of a history can be extended to truth values for all propositions, with logical properties that are to be expected from identifying truth values with probabilities.

 {\bf The Open Future} We find it hard, in a scientific theory, to accommodate the idea that there is no definite future. To be sure, we have indeterministic theories in which the future is not uniquely determined by the past, but such stochastic theories deal with complete histories encompassing past, present and future; probabilities refer to which of these histories is actual. Indeterminism, in the usual stochastic formulation, consists of the fact that there are many such histories containing a given past up to a certain time, so the future extension is not unique; but the underlying assumption is that only one of these future histories is real, so that the future is fixed even though it is not determined. In contrast, the formulation of quantum mechanics outlined here --- or what Bell \cite{Bell:cosmologists} called the ``Everett (?) theory'' --- is, I think, the only form of scientific theory in which the future is genuinely open. Unlike Bell, I do not regard this as a problem for the theory; it tells a truth which we should be happy to acknowledge. The function of the theory is to provide a catalogue of possibilities and specify how these change (deterministically) with time; it does not and cannot say which of the possibilities is actualised at any time. The ``measurement problem'' of quantum theory is no more than the difficulty of accepting this format for a scientific theory; with a change of gestalt, we can see it as a natural way to formulate indeterminism.

However, I must emphasise the roles that entanglement and the concept of internal truth play in this resolution of the measurement problem. Without these, there would be a ``preferred basis'' problem: if the universal wave function is a catalogue of possibilities, what basis defines the components which are to be regarded as possibilities? But there is no preferred-basis problem in the relative-state interpretation, as understood here. The possibilities are given by experience states, which only exist if the universal Hilbert state has a tensor product structure in which one of the factors describes a system capable of experience, i.e. which has a basis of states exhibiting the structure of propositions describing experience. It is not required that this structure should be unique; in principle, it is  possible that the universal Hilbert space has more than one tensor product structure with the required properties. If this should be so, statements about these different experiences would also be (internally) true, relative to these different structures; this would not detract from the truth of the original experience propositions. Both kinds of internal proposition would be compatible with the external truth of the same universal state vector. 

This potential ambiguity is not peculiar to quantum mechanics: a theory of conscious observers in classical physics would also admit the logical possibility that a single physical structure could admit two different interpretations in terms of conscious beings. It is hard to imagine that this could actually occur with systems complex enough to record experiences; nevertheless, it has been shown that in simpler systems such ambiguous factorisation can arise in quantum mechanics, and that decoherence can be exhibited in both factorisations \cite{tensorambig}. This does not show that an understanding based on such factorisation, like that outlined here, is untenable.


\section{Summary}

Here is how I understand nonrelativistic quantum mechanics.

\medskip

{\bf 1.} From the external perspective there is, at each time $t$, a true description of the physical 
world given by a state vector $|\Psi(t)\>$ in a Hilbert space $\H_U$.

\medskip

{\bf 2.} The sequence of states $|\Psi(t)\>$ satisfies the Schr\"odinger equation with a universal Hamiltonian $H_U$.

\medskip

{\bf 3.} In general, a description of the physical world from the external perspective is given by a closed subspace of the Hilbert space $\H_U$. Such a description has, at time $t$, the degree of truth $\<\Psi(t)|\Pi|\Psi(t)\>$ where $\Pi$ is the orthogonal projection onto the relevant closed subspace.

\medskip

{\bf 4.} Suppose the universe has a subsystem $S$ which has sufficient structure to experience and record propositions about the physical world. Then $\H_U = \H_S\otimes\H'$ where $\H_S$ is the state space of the experiencing subsystem $S$. Let $\{|\eta_n\>\}$ be an orthonormal basis of $\H_S$ which includes all possible experiences of $S$. Then the true state of the universe (from the external perspective) can be expanded as 
\[
|\Psi(t)\> = \sum_n|\eta_n\>|\Phi_n(t)\>
\]
with $|\Phi_n(t)\>\in \H'$. From the external perspective, an experience $|\eta_n\>$ is real at time $t$ if $|\Phi_n(t)\> \neq 0$, and the component $|\eta_n\>|\Phi_n(t)\>$ describes a world which has a degree of reality $\<\Phi(t)|\Phi(t)\>$ at time $t$.

\medskip

{\bf 5.} From the internal perspective of the experiencing system $S$ in the state $|\eta_N\>$ at time $t_0$, there is, for each time $t\le t_0$, one true experience (recorded in memory) described by a basis vector $|\eta_{n(t)}\>$ with $n(t_0) = N$.

\medskip

{\bf 6.} From the perspective of the experiencing system $S$ in state $|\eta_N\>$ at time $t_0$, the statement that its experience state at a future time $t > t_0$ will be $|\eta_m\>$ has truth value
\[
  \frac{\left|\<\Psi(t)|\(\Pi_m\ox I\)\e^{-iH(t-t_0)/\hbar}\(\Pi_N\ox I\)|\Psi(t_0)\right|^2} {\<\Psi(t)|\(\Pi_m\ox I\)|\Psi(t)\>\<\Psi(t_0)|\(\Pi_N\ox I\)|\Psi(t_0)\>}.
\]
where $\Pi_m$ is the orthogonal projector onto $|\eta_m\>$ in $\H_S$. The experiencing subject $S$ refers to this as the probability that they will experience $\eta_m$ at time $t$.

\medskip

{\bf 7.} From the perspective of the experiencing system $S$ in state $|\eta_N\>$ at time $t_0$, the significance of the universal state vector $|\Psi(t)\>$ is as follows.

For $t\le t_0$,  $|\Psi(t)\>$ describes what might have happened but didn't, as well as what actually did happen.

For $t > t_0$, $|\Psi(t)\>$ describes what might be going to happen at time $t$.

\medskip

Thus, from the internal perspective, the universal state vector $|\Psi(t)\>$ is not a description of reality but an influence governing changes in reality. Things that didn't actually happen still, in principle, have an effect on what is going to happen.


\end{document}